\documentclass[pdflatex,sn-mathphys-num]{sn-jnl}

\usepackage{graphicx}%
\usepackage{multirow}%
\usepackage{amsmath,amssymb,amsfonts}%
\usepackage{amsthm}%
\usepackage{mathrsfs}%
\usepackage[title]{appendix}%
\usepackage{xcolor}%
\usepackage{textcomp}%
\usepackage{manyfoot}%
\usepackage{booktabs}%
\usepackage{algorithm}%
\usepackage{algorithmicx}%
\usepackage{algpseudocode}%
\usepackage{listings}%
\usepackage{graphicx}
\usepackage{dcolumn}
\usepackage{bm}
\usepackage{flafter}
\usepackage{float}
\usepackage{lettrine}
\usepackage{amsmath}
\usepackage{amssymb}
\usepackage{color}
\usepackage{epsfig}
\graphicspath{{eps+pdf/}}
\theoremstyle{thmstyleone}%
\theoremstyle{thmstyletwo}%

\theoremstyle{thmstylethree}%

\begin{document}

\title[Article Title]{Comparative high-pressure structural and electrical transport properties  study of thermoelectric (Bi$_{1-x}$Sb$_x$)$_2$Te$_3$ compounds}

\author[1,2]{\fnm{Chenxin} \sur{Wei}}
\author[2]{\fnm{Dawod} \sur{Muhamed}}
\author[1,2]{\fnm{Wenting} \sur{Lu}}
\author[1,2]{\fnm{Haikai} \sur{Zou}}
\author[1,2]{\fnm{Baihong} \sur{Sun}}
\author[1,2]{\fnm{Shiyu} \sur{Feng}}
\author[1]{\fnm{Qian} \sur{Zhang}}
\author[2]{\fnm{Zegeng} \sur{Su}}
\author[3]{\fnm{Hirokazu} \sur{Kadobayashi}}
\author[4]{\fnm{Martin} \sur{Kunz}}
\author[5]{\fnm{Bihang} \sur{Wang}}
\author[1,6]{\fnm{Azkar, Saeed}  \sur{Ahmad}}
\author*[2]{\fnm{Yaron}  \sur{Amouya}}\email{amouyal@technion.ac.il}
\author*[1,2,6]{\fnm{Elissaios}  \sur{Stavrou}}\email{elissaios.stavrou@gtiit.edu.cn}

\affil[1]{\orgdiv{Materials Science and Engineering Department}, \orgname{Guangdong Technion-Israel Institute of Technology}, \orgaddress{\city{Shantou}, \postcode{515063}, \country{China}}}
\affil[2]{\orgdiv{Department of Materials Science and Engineering}, \orgname{Technion-Israel Institute of Technology}, \orgaddress{ \city{Haifa}, \postcode{3200003},  \country{Israel}}}
\affil[3]{\orgdiv{SPring-8/JASRI},  \orgaddress{ \street{1-1-1 Kouto}, \city{ Sayo-gun}, \postcode{ Hyogo 679-5198},  \country{Japan}}}
\affil[4]{\orgdiv{Advanced Light Source}, \orgname{Lawrence Berkeley National Laboratory}, \orgaddress{\city{Berkeley}, \postcode{94720}, \state{California}, \country{USA}}}
\affil[5]{\orgdiv{Deutsches Elektronen-Synchrotron}, \orgname{DESY}, \orgaddress{\city{Hamburgy}, \postcode{D-22603}, \country{Germany}}}
\affil[6]{\orgdiv{Guangdong Provincial Key Laboratory of Materials and Technologies for Energy Conversion}, \orgname{Guangdong Technion-Israel Institute of Technology}, \orgaddress{ \city{Shantou}, \postcode{515063}, \state{State}, \country{China}}}

\abstract{Thermoelectric (Bi$_{1-x}$Sb$_x$)$_2$Te$_3$ (BST-x) compounds with x=0.2, 0.7 and 0.9   have been studied using synchrotron angle-dispersive powder x-ray diffraction in a diamond anvil cell up to 25 GPa (at room temperature). The results clearly indicate that all compounds of this study follow a similar structural evolution with the one of pure Bi$_2$Te$_3$ and  Sb$_2$Te$_3$  under pressure.  From the comparison between the critical pressures of the corresponding phase transitions, a clear trend of increasing   critical pressure for the transition to the disordered solid-solution BCC phase was observed with the increase of Sb concentration. In the case of the BST-0.7, an extended stability of the solid-solution BCC phase up to, at least, 180 GPa was observed. Finally,   electrical transport properties measurements under pressure for BST-0.7, document  a reversible  pressure-induced metallization above 12 GPa.}


\maketitle

\section{Introduction}\label{sec1}
Bi$_2$Te$_3$-Sb$_2$Te$_3$ compounds ((Bi$_{1-X}$Sb$_x$)$_2$Te$_3$, BST-x) have been extensively studied due to their outstanding thermoelectric properties at near-room temperature ranging from 300-500K \cite{Mansouri2021,Sun2023,Mahan1998,Gayner2022}. The BST-x system exhibits full solubility. At ambient conditions, Bi$_2$Te$_3$, Sb$_2$Te$_3$ and  BST compounds adopt the rhombohedral tetradymite crystal structure (space group  $R-3m$ (166)), formed by monoatomic sublayers of Bi/Sb, and Te atoms \cite{Jacobsen2007,Manjon2013,Feutelais1993}, independently of the relative concentration of Bi and Sb. The quintuple layers of Te-Bi(Sb)-Te-Bi(Sb)-Te are stacked in a sequence along c-axis direction \cite{Nakajima1963,Hosokawa2019}.

The high-pressure studies of  Bi$_2$Te$_3$ and Sb$_2$Te$_3$ end members were motivated mainly by   the possibility of using pressure to improve their higher thermoelectric conversion efficiency \cite{Ovsyannikov2008,Khvostantsev1982,Jacobsen2012} and the increasing interest of the superconducting phase \cite{Ilina1975,Zhu2013,Zhang2012,Einaga2010}. Bi$_2$Te$_3$ and  Sb$_2$Te$_3$, exhibit, at least,  three phase transitions under pressure. Zhu $et$ $al.$ \cite{Zhu2011} clarified that Bi$_2$Te$_3$  transforms  from the rhombohedral $R-3m$ (six-fold) phase ($\alpha$-Bi$_2$Te$_3$ or phase I) to a monoclinic $C2/m$ (seven-fold) phase at 8 GPa ($\beta$-Bi$_2$Te$_3$ or phase II) and then to a $C2/c$ (eight-fold) phase at 14 GPa ($\gamma$-Bi$_2$Te$_3$ or phase III). At higher pressures, Bi$_2$Te$_3$ adopts a disorder solid-solution  body-center cubic (BCC, A2) $Im-3m$ structure (phase IV), with a 40\% and 60\% occupancy of the 2a Wyckoff Position (WP) for Bi and Te atoms, respectively \cite{Einaga2011,Zhu2011}.  Identical structural evolution was concluded for Sb$_2$Te$_3$  \cite{Ma2012,Souza2012}, albeit  with slightly different critical pressures. It is noteworthy, that the disordered BCC structure for Sb$_2$Te$_3$ appears at pressure above 20 GPa \cite{Ma2012}.

Studies on the structural evolution  of  Bi$_{1-X}$Sb$_x$Te$_3$ compounds under pressure are extremely limited. Indeed, Bai $et$ $al.$ \cite{Bai2022} studied Bi$_{0.5}$Sb$_{1.5}$Te$_3$ (BST-0.75) only  inside the pressure stability range of the ambient phase, $<$10 GPa. Jacobsen $et$ $al.$ \cite{Jacobsen2007}, studied BiSbTe$_3$ (BST-0.5) up to 18 GPa, and a phase transition was observed at $\approx$ 7 GPa. However, the indexing of the high-pressure phase was not the one concluded by more recent  studies on Bi$_2$Te$_3$ and Sb$_2$Te$_3$.

Motivated by the above, we have performed a detailed comparative $in-situ$ synchrotron angle-dispersive powder x-ray diffraction (ADXRD) study of three different BST-x compounds with x=0.2, 0.7 and 0.9, up to 20+ GPa. Our main motivation was to explore the structural evolution under pressure as a function of varying  Sb concentration. Our results document that the three BST compounds follow the same phase sequence with the pure Bi$_2$Te$_3$ and Sb$_2$Te$_3$ under pressure, without any noticeable effect of the Sb concentration on   the critical pressures for the first two phase transitions. On the other hand, it is apparent that the increase of the Sb concentration results to an increase of the critical pressure for the formation of the BCC phase.

In the case of BST-0.7, our study was extended to $\approx$ 180 GPa, with the aim of exploring the stability of the BCC phase under Mbar pressures. Surprisingly, we observe that the BCC solid-solution phase remains stable up to the highest pressure of this study. Finally, for the same BST-0.7, $in-situ$ electrical transport measurements as a function of pressure and temperature were performed. A clear indication for metallization was observed above $\approx$ 12 GPa.

\section{Experimental methods}
\subsection{Material Synthesis}
(Bi$_{1-x}$Sb$_x$)$_2$Te$_3$ (BST-x) specimens with $x$ = 0.2, 0.7 and 0.9  were synthesized by mixing commercial high-purity  Bi, Sb and Te chunks (Alfa Aesar; 99.999 \% purity) in quartz tubes according to the appropriate stoichiometric ratio. The tubes were sealed under vacuum (10$^{-5}$ Torr) and annealed at 1123 K for 8.5 h to melt the raw materials, and then cooled down to 300 K dueing  16 h \cite{Gayner2022}. The obtained ingots were ground into fine powders and sieved using a  56 $\mu$m mesh.

\subsection{ High pressure studies}
A BX-80 type diamond anvil cell (DAC) with  400 $\mu$m diameter diamond culets was used for the high-pressure experiments. A Rhenium foil pre-indented to $\approx$ 50$\mu$m and a central hole diameter of $\approx$ 130 $\mu$mwith  was used as gasket. Samples were  loaded near the center of the sample chamber, along with a ruby chip and gold  used as pressure markers, at the side. The shift of ruby  line \cite{Syassen2008} and the EOS of gold \cite{Anderson1989} were used to determine the pressure \textit{in-situ}. Neon (Ne) was used as pressure transmitting medium (PTM) which is capable to maintain hydrostatic environment up to 20 GPa \cite{Klotz2009}, and loaded with a gas loader. For the study of BST-0.7 up to 180 GPa, beveled 100 $\mu$m diameter diamond culets were used and a Rhenium foil pre-indented to $\approx$ 30$\mu$m with a hole of 40$\mu$m in diameter was used as gasket. Ne was used as the PTM.

\subsubsection {X-ray Diffraction}
X-ray diffraction (XRD) measurements at ambient conditions were performed with the  Cu K$\alpha$$_1$ ($\lambda$=1.540598\r{A}) X-ray line (Rigaku MiniFlex, Tokyo, Japan) to identify the starting  crystal structures, see Fig. S1. The results confirm that all BST specimens adopt the rhombohedral tetradymite $R-3m$ crystal structure  under ambient conditions, without any trace of impurity.

A Dectris Pilatus3 S 1M Hybrid Photon Counting detector   was used at the Advanced Light Source, Lawrence Berkeley National Laboratory, Beamline 12.2.2. The spot size of the X-ray probing beam was focused to about 10 x 10$\mu$m  using Kirkpatrick-Baez mirrors. More details on the XRD experimental setups are given in  Kunz $et$ $al.$ \cite{Kunz2005}. At SPring-8, beamline BL10XU, a Flat Panel X-ray Detector (Varex Imaging, XRD1611 CP3)  was used and the X-ray probing beam spot size was focused to approximately 10 x 10$\mu$m  using Kirkpatrick-Baez mirrors. More details on the SPring-8 XRD experimental setups are given in Kawaguchi‐Imada  $et$ $al$. \cite{KawaguchiImada2024}. At Beamline P02.2 at DESY, the X-ray probing beam were focused to a spot size of 2 x 2 $\mu$m  at the sample using Kirkpatrick-Baez mirrors and a PerkinElmer XRD 1621 flat-panel detector was used to collect the diffraction images of sample.

The integration of powder diffraction patterns to produce scattering intensity versus 2$\theta$ diagrams and initial analysis were performed using the DIOPTAS program \cite{Prescher2015}. Calculated XRD patterns were produced using the POWDER CELL program \cite{Kraus1996} for the corresponding crystal structures according to the EOSs determined experimentally in this study and assuming continuous Debye rings of uniform intensity. XRD patterns indexing has been performed using the DICVOL program \cite{Boultif2004} as implemented in the FullProf Suite. Rietveld refinements were performed using the $GSAS-II$ software \cite{Toby2013}.

\subsubsection {Electrical measurements}
In-situ high-pressure electrical resistance measurements for BST-0.7 were performed using the standard four-point probe method [R$_1$, R$_2$]. A T301 stainless steel gasket was first indented and a hole was drilled with a diameter of 300 $\mu$m. Cubic boron nitride (cBN) insulation powder was used to fill the hole and cover the steel to minimize its influence on the electrical measurement. After compressing cBN to $\approx$20 GPa, a smaller hole was drilled in the center with $\approx$150 $\mu$m in diameter. The rest of the gasket was covered with an epoxy insulating layer to further protect the electrode leads from the metallic gasket. Four electrodes were cut from a Pt foil and used for the electrical connection with the specimen.

The sample chamber was fully filled with the sample along with a small ruby chip at the center of the anvil for pressure calibration. No PTM was used during the measurement because it would permeate the space between the sample powders and interfere with their intimate contact upon pressing. The sample and the Pt probes were pressurized to about $\approx$ 2 GPa in the DAC to achieve full electrical contact of the sample with the probes. Ruby powder was used as the pressure marker

\section{Results}

\subsection{X-ray Diffraction}

\subsubsection{X-ray Diffraction of BST-0.2,0.7 and 0.9  up to 20-25 GPa }
XRD patterns of the BST specimens of this study at selected pressures are shown in Figure 1.  All BST compounds remain in the rhombohedral phase (phase I) from ambient pressure up to $\approx$ 8-10 GPa. Above this pressure, a clear phase transition was observed, based on the gradual appearance and disappearance of Bragg peaks, towards the first high pressure phase (phase II). The coexistence of phase I and phase II is apparent up to 2-3 GPa above the first indication of the appearance of phase II. Above this pressure only the peaks of phase II can be detected.  Above $\approx$ 15 GPa, another phase transition is evident towards the second high-pressure phase, phase III. As in the case of the Phase I $\rightarrow$ Phase II transition, Phases II and III also coexist for a certain pressure range. However, a clear increase of the pressure range of the co-existence is observed with increasing concentration of Sb.  At even higher pressures, new Bragg peaks appear for BST-02 and BST-09, above 17.2 and 23 GPa, respectively; see arrows in Fig. 1(a) and (c).   This is attributed to the third high-pressure phase IV that coexists with phase III for BST-0.2 and BST-0.9 up to the highest pressures of this study.

\begin{figure}[h]
{\includegraphics[width=0.9\linewidth]{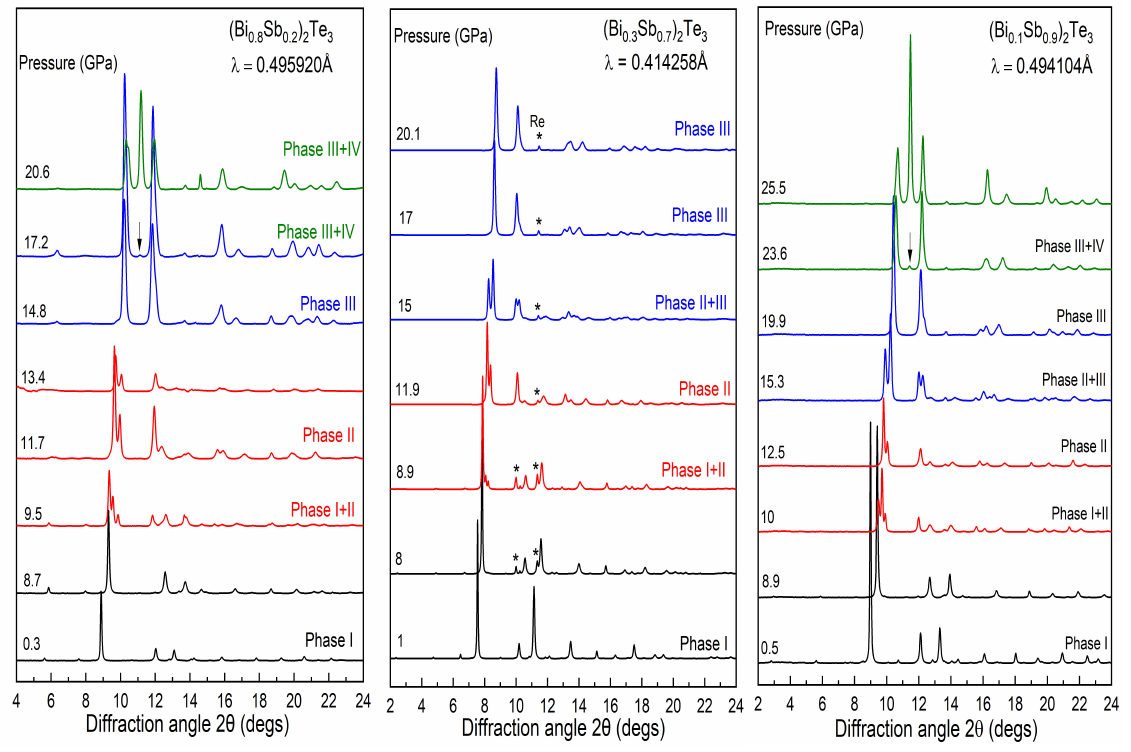}}
\centering
\caption{ XRD patterns of (a) BST-0.2, (b) BST-0.7 and (c) BST-0.9 at selected pressures. The peaks marked with asterisks originate from  Re gasket material. The relevant predominant phases, Phase I ($R-3m$), Phase II ($C2/m$) and Phase III ($C2/c)$  are noted with different colours. The arrows in (a) and (c) denote the first appearance of the strongest peak (110) of the BCC phase (phase IV). The corresponding X-ray wavelengths are noted inside the panels.}
\end{figure}

A more detailed comparison of the high-pressure phases reveals that all BST specimens adopt identical high-pressure phases: phase II and III; see Fig. S2. Moreover, both phases II and III can be indexed with the previously reported Phase II (space group  $C2/m$ (12), Z=4)  and phase III (space group  $C2/c$ (15), Z=4)  monoclinic phases of Bi$_2$Te$_3$ at similar pressures \cite{Zhu2011}. Thus, we conclude that all the BST specimens of this study qualitatively follow the same structural evolution, including the co-existence of phases pressure ranges, under pressure with Bi$_2$ Te$_3$, $i.e.$ rhombohedral $R-3m$ (phase I) $\rightarrow$ monoclinic $C2/m$ (phase II) $\rightarrow$  monoclinic $C2/c$ (phase III)$\rightarrow$ disordered BCC (phase IV). The reader is referred to the  bar diagram in Fig. 2 for a detailed account of the critical pressures and stability pressure ranges for the observed phase transitions and phases.

\begin{figure}[H]
{\includegraphics[width=0.9\linewidth]{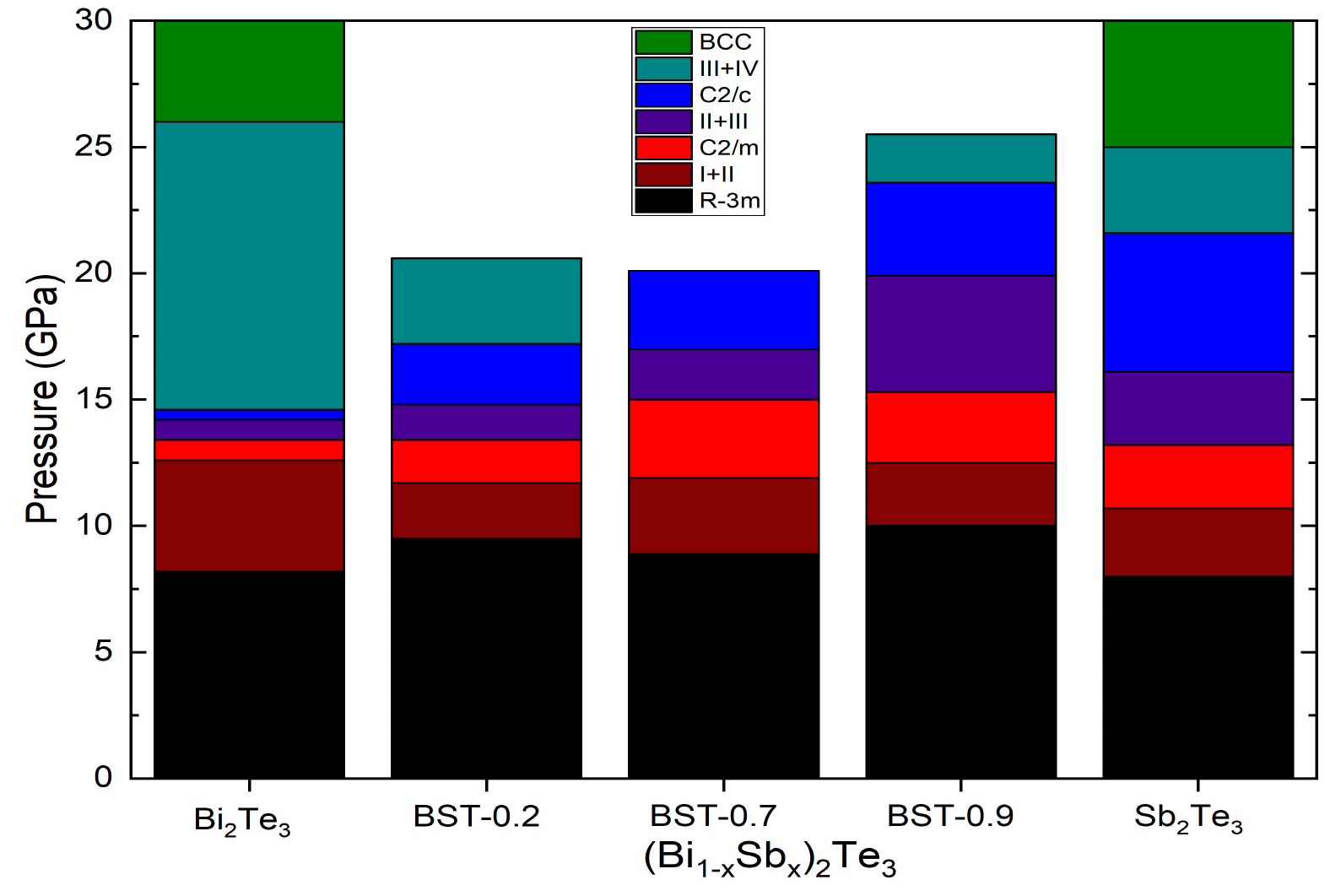}}
\centering
\caption{ Bar diagram showing the pressure stability intervals of the different structural modifications (Phase I ($R-3m$), Phase II ($C2/m$), Phase III ($C2/c)$ and phase IV ($Im-3m$) of BST compounds and Bi$_2$Te$_3$ and Sb$_2$Te$_3$ end members. The data for  Bi$_2$Te$_3$ and Sb$_2$Te$_3$ are adopted from Ref. \cite{Zhu2011} and Ref. \cite{Ma2012}, respectively.}
\end{figure}

From the relevant Rietveld refinements of the XRD patterns, see Fig. S3, using the crystal structures mentioned above, the experimental lattice parameters and cell volumes were determined and are plotted as a function of pressure in Fig. 3 and listed in Table 1. From the data of Fig. 3(d)-(f), the bulk modulus \textit{B} and its pressure derivative \textit{B’} were determined by fitting a third-order Birch-Murnaghan \cite{Birch1978} Equation of State (EOS), Eq. 1, to the experimental data, see Table 1.

\begin{equation}
\scriptsize
P = \frac{3}{2} B_{T_0} \left[ \left( \frac{V_0}{V} \right)^{\frac{7}{3}} - \left( \frac{V_0}{V} \right)^{\frac{5}{3}} \right]
\times \left\{ 1 + \frac{3}{4} \left( B'_{T_0} - 4 \right) \left[ \left( \frac{V_0}{V} \right)^{\frac{2}{3}} - 1 \right] \right\}
\end{equation}

\begin{table}[h]
\centering
\scriptsize
 \caption{Experimental structural parameters of BST-0.2, 0.7 and 0.9 at selected pressures: space group (SG), number of formula units in the unit cell Z, lattice parameters, cell volume per formula unit V$_{pfu}$, bulk modulus \textit{B} and its pressure derivative \textit{B'} (as determined by fitting a 3$^{rd}$ order Birch-Murnaghan  EOS \cite{Birch1978}  to the  experimental data) at the  onset pressure.}
    \begin{tabular}{ccccccccc}
         P (GPa)&  SG&  Z&  a (\r{A})&  b (\r{A})&  c(\r{A})&  V$_{pfu}$ (\r{A}$^3$)&  B (GPa)& B'\\ \hline
         &&&&&BST-0.2&&&\\\hline
         ambient&  R-3m&  3&  4.359& 4.359&  30.506&  167.316&  39(4)& 6.0 (0.4)\\
         0.9&  R-3m&  3&  4.351& 4.351&  30.341&  165.797&  & \\\hline
         10.5&  C2/m&  4&  14.688&  4.141&  8.958&  136.151&  79.4 (4.9)& 4 (fix)\\\hline
         14.8&  C2/c&  4&  9.991&  6.918&  7.706&  125.471&  82.5 (4.6)& 4 (fix)\\ \hline
         &&&&&BST-0.7&&&\\\hline
         ambient&  R-3m&  3&  4.303& 4.30 &  30.430&  162.638&  22.9 (10)& 13.5 (1.1)\\
         3&  R-3m&  3&  4.206& 4.206 &  29.563&  150.981&  & \\\hline
         14&  C2/m&  4&  14.361&  4.007&  8.781&  126.306&  95 (6)& 4 (fix)\\\hline
         20.1&  C2/c&  4&  9.692&  6.737&  7.574&  116.181&  125.3 (1.6)& 4 (fix)\\ \hline
         &&&&&BST-0.9&&&\\\hline
         ambient&  R-3m&  3&  4.272& 4.272&  30.390&  160.701&  29.1 (2.0)& 8.7 (1.2)\\
         0.5&  R-3m&  3&  4.260& 4.260&  30.264&  158.551&  & \\\hline
         12.5&  C2/m&  4&  14.319&  3.974&  8.879&  126.269&  69 (8)& 4 (fix)\\\hline
         19.9&  C2/c&  4&  9.749&  6.738&  7.440&  114.632&  147 (11)& 4 (fix)\\
\end{tabular}
\end{table}
%\onecolumngrid

\begin{figure}[H]
{\includegraphics[width=\linewidth]{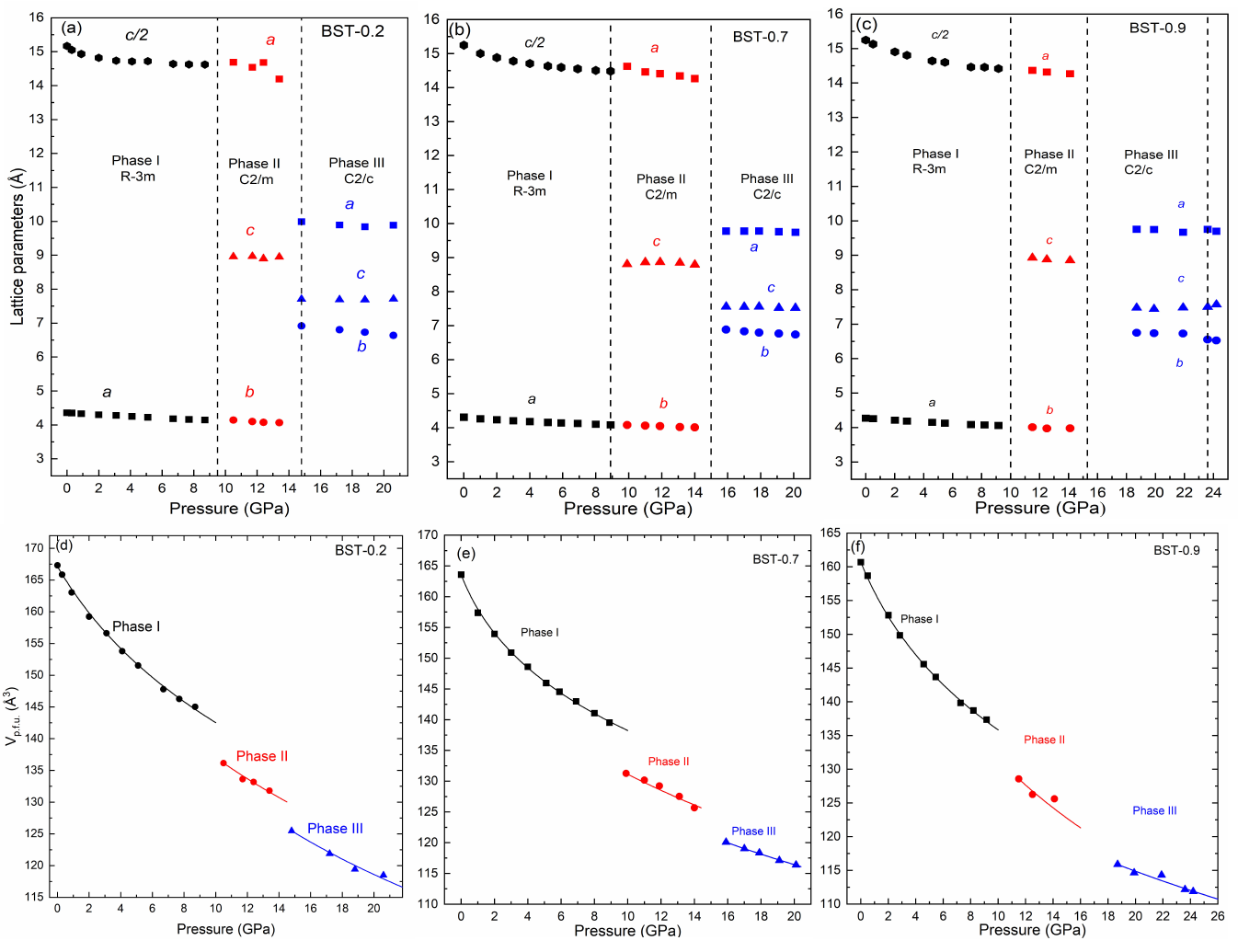}}
\centering
\caption{ Pressure dependence of the lattice parameters and cell volume per formula unit of (a),(d) BST-0.2, (b),(e) BST-0.7 and (c),(f) BST-0.9 as a function of pressure.  The vertical dashed line indicates the critical pressures of the phase transitions.  }
\end{figure}

%\twocolumngrid

\subsubsection{X-ray Diffraction of BST-0.7  up to Mbar pressures}
Aiming to explore the pressure stability range of the disordered solid-solution BCC phase, BST-0.7 was pressurized up to 180 GPa. Part of our motivation was to explore the effect of solid-solution, as opposed to the ordered B2 CsCl-type phase, to the pressure stability of the BCC phase. The B2 phase of NaCl  remains stable up to $\geq$300 GPa \cite{Ono2010,Sakai2011}, while it is predicted to transform to an orthorhombic oC8 phase above 325 GPa \cite{Chen2012}. On the other hand, the much heavier elements CsI, that adopts the CsCl-type structure under ambient conditions, has been reported to transform to an orthorhombic structure already at 45 GPa, and finally transform to an HCP-like structure at higher pressures \cite{Mao1989,Mao1990}.

\begin{figure}[H]
{\includegraphics[width=0.9\linewidth]{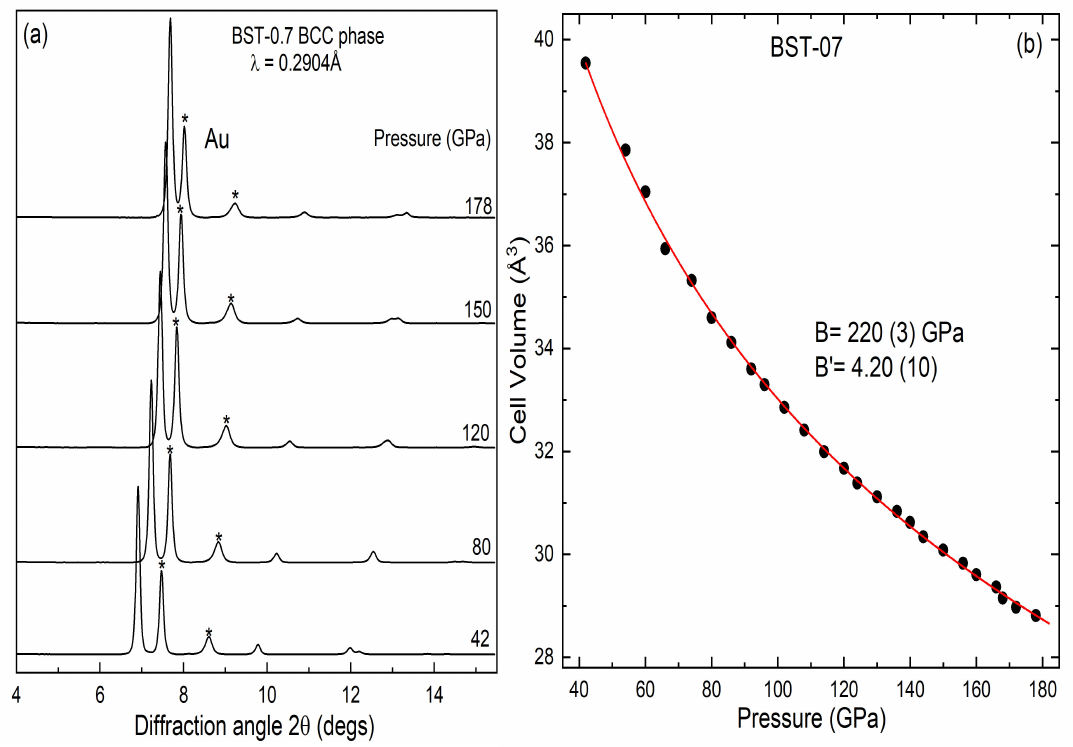}}
\centering
\caption{ (a) XRD patterns of   BST-0.7  at selected pressures up to 178 GPa.The asterisks denote the Bragg peaks from Au (pressure marker). (b) Pressure dependence of the BCC cell volume of  BST-0.7. The red solid curve is a 3$^{rd}$ order Birch-Murnaghan  EOS \cite{Birch1978}  fitted to the  experimental data. The determined bulk moduli at the onset pressure are also given.  }
\end{figure}

The XRD results of this study indicate that BST-0.7 remains in the BCC phase up to, at least, 180 GPa, see Fig. 4. For comparison, from previous studies it has been established  that Bi remains in the BCC phase up to at least 220 GPa \cite{Akahama2002}. On the other hand, the BCC phase of Te transforms  to the FCC phase at 99 GPa, and the FCC phase remains stable up to, at least, 330 GPa \cite{Akahama2018}. Finally, Sb adopts  a BCC structure at 28 GPa, that remain stable up to 43 GPa \cite{Degtyareva2004}. According to Ref. \cite{Zhu2011}, pressure results to a substantial Bi$\rightarrow$Te charge transfers, resulting in an enhanced HP ionicity of the BCC phase of Bi$_2$Te$_3$. Thus, the extended stability of the BCC phase could presumably be attributed to this enhanced ionicity that stabilizes the BCC phase. A similar scenario was determined for the extended stability of the $\alpha$-Mn phase, above the pressure that the HCP phase becomes energetically favorable \cite{MagadWeiss2021}.

\subsection{Electrical measurements}
The  electrical resistance of BST-0.7 as a function of pressure, at room temperature, is shown in Fig. 5(a), for both the compression and decompression runs. Our main motivation for studying the electrical properties of BST-0.7 under pressure was our observation that, at ambient pressure, BST-0.7 shows an almost temperature-independent conductivity, in contrast to the clear semiconducting and metallic behavior of BST-0 (Bi$_2$Te$_3$) and BST-1 (Sb$_2$Te$_3$), respectively, see Fig. S4 \cite{Dawod2025}.  Upon initial compression, the resistance rapidly decreases, reaching an almost constant value above $\approx$ 11-12 GPa. Such a pressure-dependent behavior of the  resistance  is attributed to a pressure-induced semiconductor to metal transition \cite{Lu2025}. A similar pressure  dependence of the electrical resistivity was also observed in the case of pure Bi$_2$Te$_3$ \cite{Zhang2013}. Upon decompression, the resistance  starts to increase around 6 GPa, and reaches almost the same, with initial compression, value  at near ambient pressure with a  hysteresis of $\approx$ 5-6 GPa.

\begin{figure}[h]
{\includegraphics[width=0.8\linewidth]{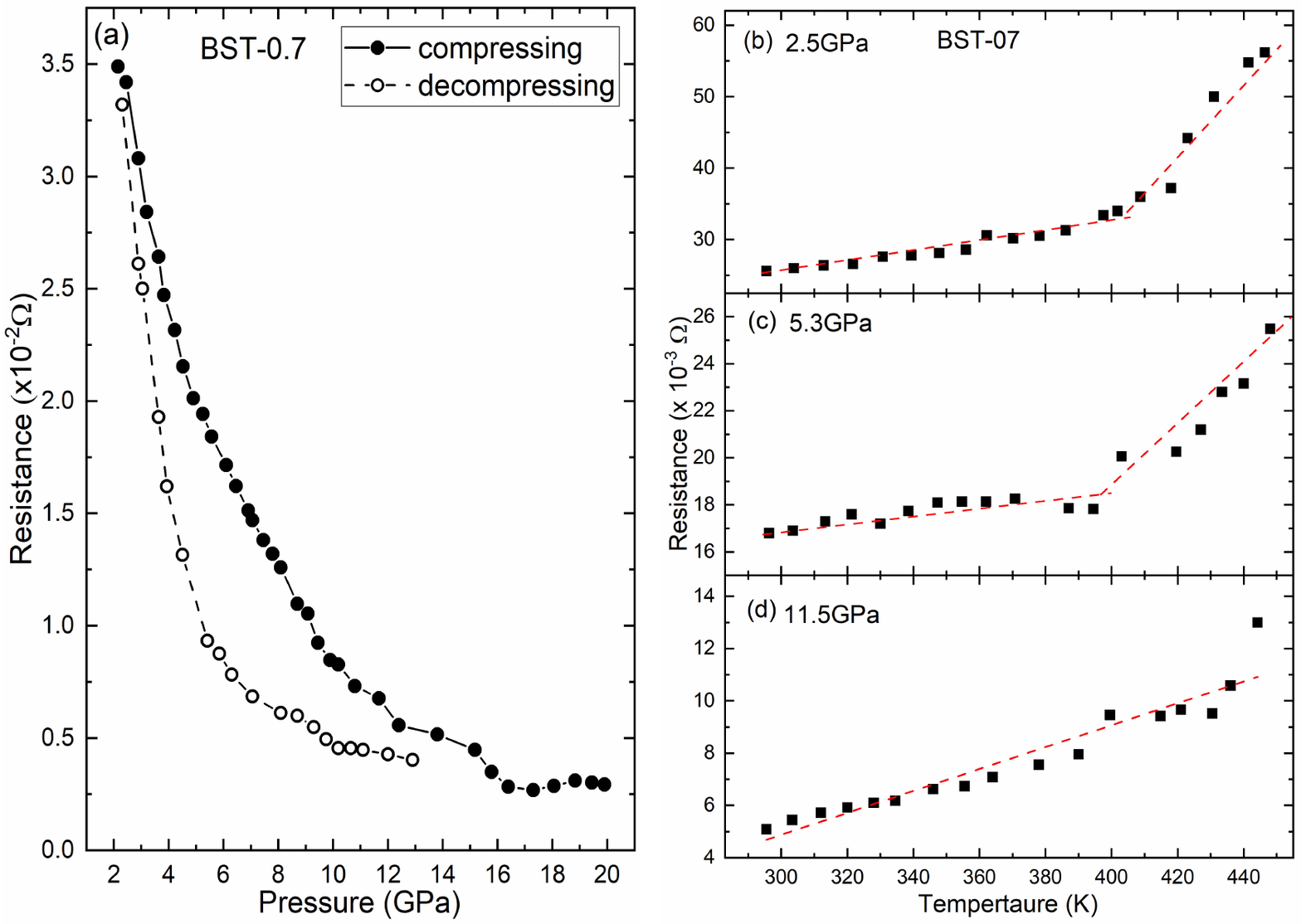}}
\centering
\caption{ (a) Electrical resistance of BST-0.7 as a function of pressure, at room temperature, under compression (solid symbols) and decompression (open symbols). Electrical resistance of BST-0.7 as a function of temperature at (b) 2.5 GPa, (c) 5.3 GPa and (d) 11.5 GPa. The dashed red lines are guides to the eye, see text for details.}
\end{figure}

To confirm the pressure-induced metallization, we performed high-temperature electrical measurements at different pressures, see Figs. 5(b)-(d). We note that, because of different experimental runs, and the fact that resistance is not an intrinsic property of the sample,  comparison between absolute resistance values between runs is challenging. However, the temperature dependence of the resistance (metallic character) is of major importance.  At 2.5 and 5.3 GPa, the resistance remains practically constant up to a certain temperature, indicative of a material with very narrow band-gap, and then linearly increases with temperature,  thus showing a clear metallic character. On the other hand, at 11.5 GPa, the resistance increases monotonically with temperature, highlighting the metallic character of the material at this pressure from room temperature.  We note that, although challenging to establish a definite crossover temperature, the transition from narrow-bang-gap $\rightarrow$ metallic character appears to occur at lower T for 5.3 GPa than 2.5 GPa; see dashed red lines in Figs. 5(b) and (c).

\section{Discussion }
\subsubsection{ Pressure dependence of the $c/a$ axial ratio in the ambient phase of BST-$x$ }

In previous high-pressure studies of the ambient phase of Bi$_2$Te$_3$, a crossover of the pressure dependence of the  $c/a$ axial ratio was observed \cite{Polian2011,Nakayama2009}. In details, $c/a$ initially decreases, as expected based on weaker van der Waals interlayer ($c$-axis)  bonding, followed by an increase above 2.5-3.5 GPa. Polian $et al$. \cite{Polian2011},  attributed this trend to an electronic topological transition (ETT), that mainly affects the intralayer electronic distribution  of the strong ionocovalent bonds,  rendering $a$-axis more compressible. The same crossover behavior of the $c/a$ ratio was observed for all the BST compounds in this study, see Fig. 6. Although determination of the exact crossover pressure is challenging, no clear trend was observed with varying Sb concentration. Thus, we conclude that the appearance of the EET is universal for BST compounds and  independent of the Sb concentration.

\begin{figure}[h]
{\includegraphics[width=0.5\linewidth]{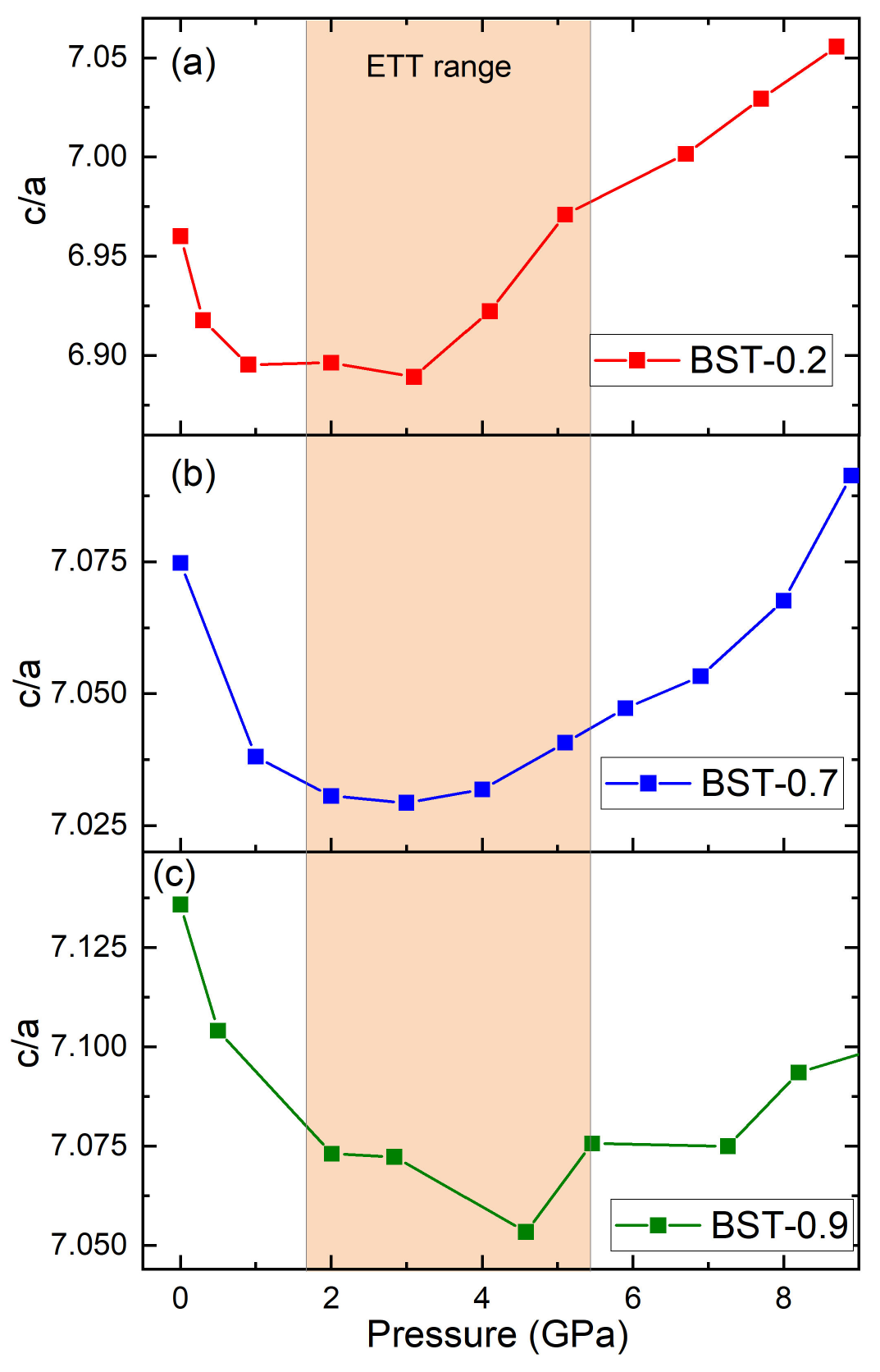}}
\centering
\caption{$c/a$ ratios of the $R-3m$ phase of a) BST-0.2, b) BST-0.7 and c) BST-0.9 as function of pressure. The orange rectangle denotes the range of the EET transition, see text. The solid lines are guides for the eye. }
\end{figure}

\subsubsection{Effect of the Sb concentration on the critical pressure for the formation of the solid-solution BCC phase }

The structural analysis of the BST compounds in this study clearly indicates  the critical pressure for the formation of the BCC phase increases with increasing Sb concentration. In detail, for BST-0.2 the first sign for the appearance of the BCC phase is at 17.2 GPa, no BCC phase was observed for BST-0.7 up to 20 GPa, and for BST-0.9 BCC appears at 23.6 GPa, see Figs. 1 and 2. The observed trend is in agreement with that reported  for pure Bi$_2$Te$_3$ (BST-0) \cite{Zhu2011, Einaga2011} and the corresponding observation of the BCC phase above 14.5 GPa. A previous study on  Sb$_2$Te$_3$, with the appearance of BCC phase above 21.6 GPa \cite{Ma2012}, seems to contradict this trend. However, this  apparent disagreement can be resolved considering that in Ref. \cite{Ma2012}, a 4:1 methanol-ethanol mixture was used as PTM, that is heavily non-hydrostatic above 12 GPa \cite{Klotz2009}. It is well known that non-hydrostatic conditions promote phase transitions to occur at lower pressures \cite{Zhang2023}. A similar trend could also be claimed for the formation of the $C2/c$ phase, see Fig.2. However, the relatively small pressure differences and experimental limitations preclude a conclusion about a definite trend.

The trend discussed above, of increasing critical pressure for the formation of a solid-solution BCC phase upon increasing the Sb concentration, appears to be counterintuitive.  Under ambient conditions, the atomic radius of Sb is closer (than that of Bi) to Te, and according to theoretical calculations, this continues to be valid under pressure \cite{Rahm2016,Rahm2020}. Thus, and having in mind the Hume-Rothery rule that solid-solutions are promoted when atomic radii are close,  the opposite effect shall be expected. To conclude, the underlying reason for the observed trend is still not well understood, and further studies, presumably theoretical, are needed to elucidate this effect.

\subsubsection{Electrical properties under pressure}
As mentioned in the results section, the transition from narrow-bang-gap $\rightarrow$ metallic character appears to occur at lower T upon pressure increase. This is in agreement with the general intuition that pressure promotes metallic character. Moreover, previous density functional theory (DFT) calculations \cite{Gayner2022} for the ambient conditions $R-3m$ phase of  Bi$_2$Te$_3$, document that application of compressive strain along the $c$-axis results to an abrupt  decrease of the band gap, eventually reaching metallization at $\approx$12 \% compressive strain. Although a direct comparison with the present results under hydrostatic pressure is not possible,  BST-0.7 undergoes a phase transition when the $c$-axis relative compression becomes higher than $\approx$5 \% (see Fig. 3(b)) at $\approx$ 9 GPa, a common trend can be inferred. Nevertheless, the previous DFT calculations and present results do agree  on the rapid decrease of the bang gap under compression.

Finally, we note that although attributing the crossover of the pressure dependence of the  $c/a$ axial ratio to an EET transition looks reasonable, from our electrical measurements no direct effect on the electrical conductivity as a function of pressure was observed at the relevant pressure range. Indeed, the electrical conductivity shows a smooth variation between 2-8 GPa.

\section{Conclusions}
All the  BST-x compounds (x=0.2, 0.7 and 0.9) investigated in this study exhibit identical structural evolution under pressure, that is also identical with the one for pure  Bi$_2$Te$_3$ and  Sb$_2$Te$_3$. Sb concentration marginally, if at all, affects critical pressures for the  phase I $\rightarrow$ phase II $\rightarrow$  phase III phase transitions. However, increasing Sb concentration results to a clear increase of the critical pressure for the transition to the disordered BCC phase IV.  The exact origin of this effect is still not clear. In the case of BST-0.7, the BCC phase was found to have an extensive pressure stability up to 180 GPa. For the same BST alloy, a pressure induced  metallization was observed above  12 GPa. Finally, a clear crossover of the value of the $c/a$ axial ratio of the ambient phase with increasing pressure, previously attributed to a pressure-induced  EET transition, was observed for all BST compounds of this study.

\backmatter

\bmhead{Supplementary information}
Supplementary Figs. 1-4 are provided in Supplementary information file.

\bmhead{Acknowledgements}
C.W. and S.F. acknowledge support from the Graduate Scholarships of the Guangdong Provincial Key Laboratory of Materials and Technologies for Energy Conversion. Y.A. would like to acknowledge generous support from the Pazy Research Foundation, Grant No. 2032063. The work performed at GTIIT was supported by funding from the Guangdong Technion Israel Institute of Technology and the Guangdong Provincial Key Laboratory of Materials and Technologies for Energy Conversion, MATEC (No. MATEC2022KF001). Beamline 12.2.2 at the Advanced Light Source is a DOE Office of Science User Facility under contract no. DE-AC02-05CH11231.Part of the synchrotron radiation experiments were performed at BL10XU  of SPring-8 with the approval of the Japan Synchrotron Radiation Research Institute (JASRI) (Proposal Nos. 2024A1096 and 2024B1182). We also acknowledge DESY (Hamburg, Germany), a member of the Helmholtz Association HGF, for the provision of experimental facilities. Parts of this research were carried out at PETRA III beamline P02.2.

\subsection*{Conflict of Interest}
The authors have no conflicts to disclose.

\subsection*{Data Availability}
The data that support the findings of this study are available from the corresponding author upon reasonable request.

\end{document}